\documentclass[%
 reprint,
 amsmath,amssymb,
 aps,
]{revtex4-1}

\usepackage{graphicx}
\usepackage{dcolumn}
\usepackage{braket}
\usepackage{gensymb}
\usepackage{bm}
\usepackage{amsmath}
\usepackage{float}
\usepackage{color}
\usepackage[utf8]{inputenc}

\begin{document}

\preprint{APS/123-QED}

\title{Quantum Counterfactual Communication Without a Weak Trace}
\author{D. R. M. Arvidsson-Shukur}
\author{C. H. W. Barnes}%
\affiliation{%
 Cavendish Laboratory, Department of Physics, University of Cambridge, Cambridge CB3 0HE, United Kingdom
}%

\date{\today}

\begin{abstract}
The classical theories of communication rely on the assumption that there has to be a flow of particles from Bob to Alice in order for him to send a message to her. We develop a quantum protocol that allows Alice to perceive Bob's message ``counterfactually". That is, without Alice receiving any particles that have interacted with Bob. By utilising a setup built on results from interaction-free measurements, we outline a communication protocol whereby the information travels in the opposite direction of the emitted particles. In comparison to previous attempts on such protocols, this one is such that a weak measurement at the message source would not leave a weak trace that could be detected by Alice's receiver. Whilst some interaction-free schemes require a large number of carefully aligned beam-splitters, our protocol is realisable with two or more beam-splitters. We demonstrate this protocol by numerically solving the time-dependent Schr\"odinger Equation (TDSE) for a Hamiltonian that implements this quantum counterfactual phenomenon.
\begin{description}
\item[PACS numbers]
03.65.Ta, 03.67.Hk, 03.67.Ac
\end{description}
\end{abstract}

\pacs{Valid PACS appear here}
\maketitle


\section{Introduction} A century ago, the discovery of quantum mechanics caused a renaissance of physics as a subject. The view of the fundamental nature of physical phenomena was drastically changed. During the following century the scientific community saw several ideas put forward of how to manifest the novelties of quantum mechanics.\cite{Bell1964a, Hardy92} Furthermore, the novelties of quantum mechanics led to the development of quantum technologies, which can provide solutions to problems that classical systems cannot solve.\cite{Wiesner83, Feynman82,  Bennett84, Ekert91, Bennett93, Knill01} 

A physical novelty that quantum mechanics provide, is the interaction-free measurement (IFM), first developed by Elitzur and Vaidman.\cite{Elitzur93} An IFM uses a probing interrogating particle sent through a quantum self-interference device, such as a Mach-Zehnder Interferometer (MZI), in order to obtain information about whether or not an object exists at a certain location. By utilising the postulate of wavefunction collapse \cite{Nielsen11}, the protocol can be carried out such that the interrogating particle never directly interacts with or is deflected by the object of interest. These types of non-interacting interrogations are also referred to as counterfactual processes.\cite{Hosten06, Salih13, Li15}

Significant improvements to classical information theory have also been attributed to quantum mechanics. Classically, Shannon showed how many bits have to travel from Bob to Alice in order for Bob to provide Alice with a message containing certain information.\cite{Shannon48} The classical assumption that one bit of information had to be carried by a one bit particle, was challenged by the quantum concept of superdense coding, put forward by Bennett and Wiesner in 1992. Superdense coding allows Bob to send two classical bits encoded in only one quantum particle (qubit).\cite{Bennett92} Schumacher then extended many ideas from classical information theory to the quantum mechanical scenario, showing how classical information can be efficiently encoded in qubit particles and sent from Bob to Alice over a quantum channel.\cite{Schumacher95}

\color{black}
Moreover, quantum information theory led to the development of unconditionally secure quantum key distribution (QKD) schemes.\cite{Bennett84, Ekert91, Bennett92, Shor00, Quan02} It was later shown \cite{Noh09} that the distribution process of secret keys in QKD protocols can be realised with counterfactual phenomena, without the secret key particles ever traveling between the communicating parties. Such schemes have been experimentally realised \cite{Ren11, Brida12, Liu12}, and their security advantages over other QKD protocols have been studied \cite{Yin10, Yin12, Liu14}. Whilst the secret key is generated counterfactually, the classical public channel communication of these schemes requires particle transfer.
\color{black}

Classically, the exchange of physical particles in the direction of the message has been assumed necessary for information transfer between two communicating parties, Alice and Bob.\cite{Shannon48} However, could it be possible that quantum mechanics enables counterfactual transfer (without any particle exchange) of messages from Bob to Alice?

Salih et al. have previously \cite{Salih13} attempted to produce methods for such counterfactual communication, using schemes similar to those presented in Ref. \cite{Hosten06}. These methods crucially depend on nested MZI devices.  However, such devices have been the subject of intense debate in recent years \cite{Vaidman13, Li13, Vaidman13-2, Salih13, Vaidman14, Salih14, Li15, Vaidman16, Li16}. The criticism put forward---primarily by Vaidman---highlights the dilemma of \textit{welcher Weg} (``which path") determination of quantum particles. By exploring the ``weak trace" \cite{Vaidman13}---introduced by weak measurements in different spatial locations of the protocol---Vaidman shows how significant parts of wavepackets actually travel from Bob to Alice in the protocol of Salih et al.\cite{Vaidman13-2, Vaidman14}

In this paper we adopt the role of quantum diplomats and present a novel quantum mechanical communication protocol, which avoids the weak trace implications of previous works. We outline how the theory of interaction-free measurements can be utilised to create a protocol for direct information transfer. Whilst our protocol does not alter the limits of bit transfer outlined by Shannon, Bennett, Weisner and Schumacher, it does significantly alter the role of the physical particles in the transmission scheme. Our protocol contradicts the intuitive idea and tenet of information theory, that the particles that carry a message of information ought to travel in the same direction as the message. We introduce a novel concept of weak-trace-free counterfactual communication, which allows for the non-local acquiring of information about distant systems. Furthermore, we present a numerical demonstration of the protocol by solving the time-dependent Schr\"odinger equation (TDSE) of a tailored massive particle Hamiltonian.

\section{Background}
The methods presented by Salih et al. make use of a complicated IFM device, constructed from stacked nested MZIs.\cite{Salih13} The nature of their protocol presents two problems for the realisation of counterfactual communication.  Firstly, owing to the structure of the nested IFM device, their protocol fails to eliminate the weak-trace impact in the laboratory of Bob (the sender).\cite{Vaidman88, Vaidman13, Vaidman14} \color{black} Secondly, owing to the experimental difficulties with the realisation of high-efficiency interaction-free measurements,\cite{Kwiat99, Namekata06, Peise15} and the fact that a success probability of $>95\%$ relies on the perfect alignment of over $60000$ beam-splitters with precise transmission and reflection coefficients,\cite{Salih13} we deem the high-fidelity physical implementation of their protocol unlikely. \color{black} 

\begin{figure}
\centering
\includegraphics[scale=0.16]{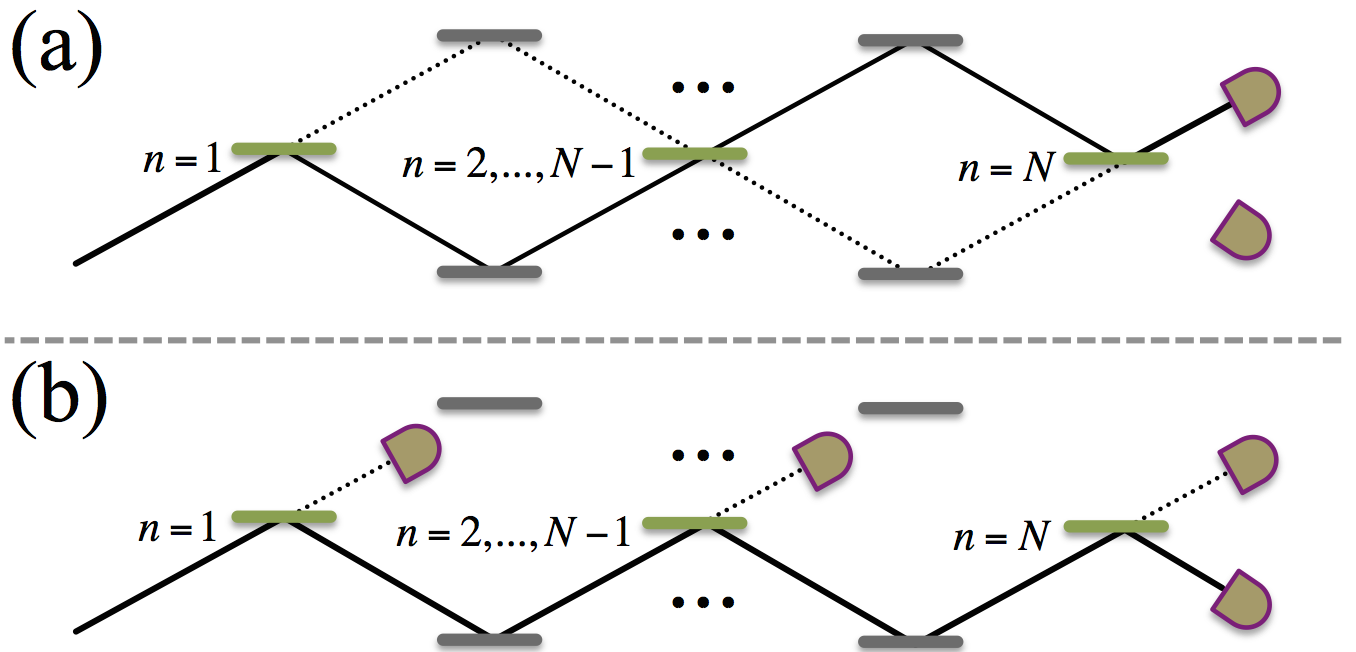}
\caption{(color online) A stacked MZI IFM device with $N$ beam-splitters. If the upper path is free (a) the particle always exits through the upper slot. If the upper path is blocked by detectors (b) the particle exits through the lower path with probability $\cos{(\pi/2N)}^{2N}$.}
\label{fig:Kwiat}
\end{figure}

In order to avoid the implications of the stacked nested IFMs for interaction-freeness and experimental feasibility, we seek to make use of stacked non-nested MZI devices, as originally developed by Kwiat et al. \cite{Kwiat95} (shown in FIG. \ref{fig:Kwiat}). Whether or not the particle will be detected at the lower or upper output is determined by the number of beam-splitters, $N$ (with reflection coefficient $\cos{\big( \frac{\pi}{2N}\big)}$), and whether or not there are detectors present in the upper path. In the scenario of $N$ perfect beam-splitters, the particle will end up in the upper output path with probability $1$ if the path is free, and in the lower path with probability $\cos{\big( \frac{\pi}{2N}\big)}^{2N}$ ($1$ in the limit of large $N$) if the upper path is blocked by detectors. If the lower path was assigned to Alice's Laboratory and the upper path to Bob's, the process with detectors present would generate a counterfactual detection for Alice. However, the scenario of an empty upper path would not, as the wavefunction of the particle would travel back and forth between the two laboratories. Nevertheless, we shall see that it is possible to avoid the exchange of wavefunction from Bob to Alice by a clever spatial arrangement of the Transmission Line and Alice's and Bob's respective laboratories (see FIG. \ref{fig:HilbSpa}).

We introduce the bosonic creation and annihilation operators of the respective spatial domains: $a^{\dagger}_{\rm{A},\rm{Tr},\rm{B}}$ and $a_{\rm{A},\rm{Tr},\rm{B}}$. The basis states of a restricted one-particle system can then be expressed as:
\begin{align}
& \ket{0}^{(\rm{A})}\ket{0}^{(\rm{Tr})}\ket{0}^{(\rm{B})}  \nonumber \\ 
a^{\dagger}_{\rm{(A)}} \ket{0}^{(\rm{A})}\ket{0}^{(\rm{Tr})}\ket{0}^{(\rm{B})}  \nonumber  = & \ket{1}^{(\rm{A})}\ket{0}^{(\rm{Tr})}\ket{0}^{(\rm{B})}  \nonumber \\  
a^{\dagger}_{\rm{(Tr)}} \ket{0}^{(\rm{A})}\ket{0}^{(\rm{Tr})}\ket{0}^{(\rm{B})}  \nonumber  = & \ket{0}^{(\rm{A})}\ket{1}^{(\rm{Tr})}\ket{0}^{(\rm{B})}  \nonumber \\  
a^{\dagger}_{\rm{(B)}} \ket{0}^{(\rm{A})}\ket{0}^{(\rm{Tr})}\ket{0}^{(\rm{B})}  \nonumber  = & \ket{0}^{(\rm{A})}\ket{0}^{(\rm{Tr})}\ket{1}^{(\rm{B})} .  
\end{align}
We also introduce a number of beam-splitters, $N$, that act between the Transmission Line and Bob's Laboratory. They implement a transformation such that:
\begin{align}
\begin{pmatrix}
   a'_{\rm{Tr}}  \\
  a'_{\rm{B}}
 \end{pmatrix}
=
\begin{pmatrix}
   \cos{(N \theta)} & i\sin{(N \theta)} \\
  i\sin{(N \theta)} & \cos{(N \theta)}
 \end{pmatrix} 
 \begin{pmatrix}
   a_{\rm{B}}  \\
  a_{\rm{Tr}}
 \end{pmatrix} ,
\end{align}
where the primed and un-primed operators denote the output and input operators respectively.

\section{The Counterfactual Communication Protocol}
\begin{figure}
\centering
\includegraphics[scale=0.17]{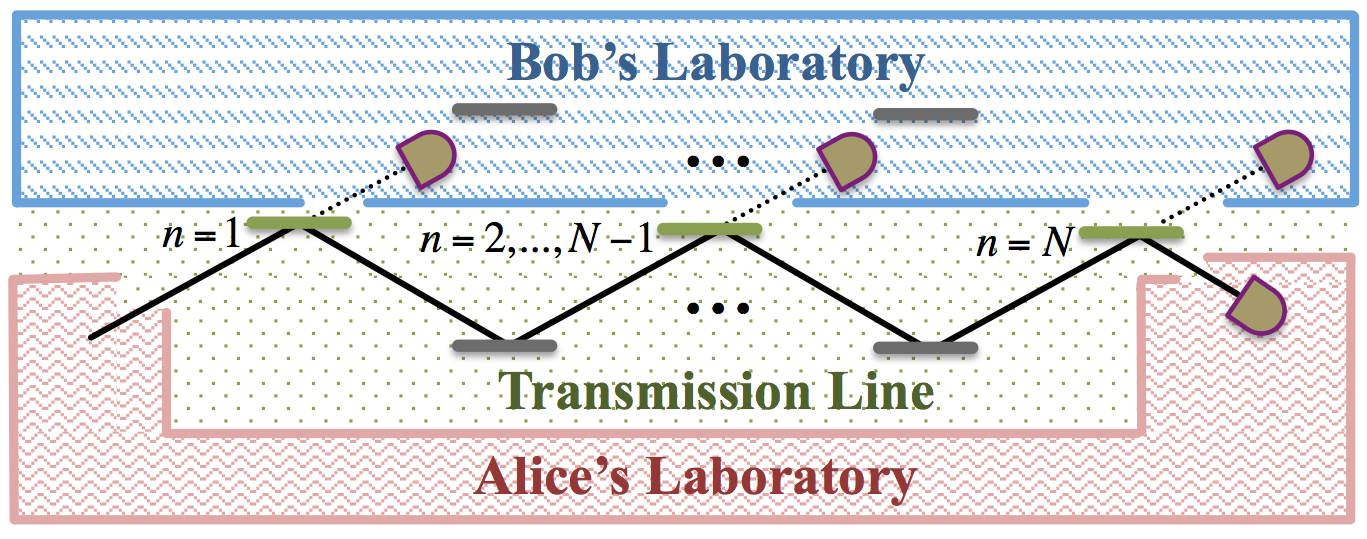}
\caption{(color online) Optical setup of a stacked IFM device showing Alice's, Bob's and the Transmission Line's spatial occupation.}
\label{fig:HilbSpa}
\end{figure}
In this section we outline the binary quantum counterfactual direct communication protocol of this paper. 

Firstly, we denote the respective Hilbert spaces of Alice's, Bob's and the Transmission Line's spatial occupation as $\mathcal{H}^{(\rm{A})}$, $\mathcal{H}^{(\rm{B})}$ and $\mathcal{H}^{(\rm{Tr})}$, such that the total Hilbert space becomes $\mathcal{H}^{(\rm{A})}  \oplus \mathcal{H}^{(\rm{Tr})} \oplus \mathcal{H}^{(\rm{B})}$. Each of these Hilbert spaces have the occupation number as their degree of freedom. Alice's Hilbert space, $\mathcal{H}^{(\rm{A})}$, contains the lower output and input ports of the stacked IFM. Bob's space, $\mathcal{H}^{(\rm{B})}$, contains the upper path of the IFM device and the upper output port. Finally, The Transmission Line's space, $\mathcal{H}^{(\rm{Tr})}$, contains the lower part of the IFM device and all the beam-splitters. See Fig. \ref{fig:HilbSpa}.

In the following two subsections we outline the two processes that make up the communication protocol. In both bit processes Bob and Alice have pre-determined time-intervals during which Alice is to make particles available to the Transmission Line. Moreover, below we omit the prime-notation as it simply indicates smaller subspaces of the defined Hilbert spaces.

\textit{(a) The $0$-Bit Process:}

\textbf{Step $\bm{1}$:} The protocol starts with Alice creating a particle in the initial vacuum state:
\begin{align}
\ket{0}^{(\rm{A})}\ket{0}^{(\rm{Tr})}\ket{0}^{(\rm{B})} \rightarrow & a^\dagger_{(\rm{A})} \ket{0}^{(\rm{A})}\ket{0}^{(\rm{Tr})}\ket{0}^{(\rm{B})} \nonumber \\ 
 & = \ket{1}^{(\rm{A})}\ket{0}^{(\rm{Tr})}\ket{0}^{(\rm{B})} .
\end{align}

\textbf{Step $\bm{2}$:} Alice sends her particle to the Transmission Line with the pre-determined frequency:
\begin{align} 
& a^\dagger_{(\rm{Tr})} a_{(\rm{A})} \ket{1}^{(\rm{A})}\ket{0}^{(\rm{Tr})}\ket{0}^{(\rm{B})} \nonumber \\ 
 & = \ket{0}^{(\rm{A})}\ket{1}^{(\rm{Tr})}\ket{0}^{(\rm{B})} .
\label{eq:aftAl}
\end{align} 

\textbf{Step $\bm{3_0}$:} If Bob wishes to transmit a $0$-bit, he makes sure that there are no detectors in the upper paths of the IFM device, i.e. in his laboratory. After the particle has entered the Transmission Line it hits a beam-splitter, after which some of the wavefunction travels to Bob's Laboratory. The beam-splitter angle is set to $\theta = \pi / 2 N$. The wavepacket falls on the beam-splitter $N$ times and the following evolution takes place:
\begin{align*}
 & a^\dagger_{(\rm{Tr})} \ket{0}^{(\rm{A})}\ket{0}^{(\rm{Tr})}\ket{0}^{(\rm{B})} \xrightarrow{BS_N}  ia^{\dagger}_{(\rm{B})} \ket{0}^{(\rm{A})}\ket{0}^{(\rm{Tr})}\ket{0}^{(\rm{B})}  \\ 
 & =i \ket{0}^{(\rm{A})}\ket{0}^{(\rm{Tr})}\ket{1}^{(\rm{B})} .
\end{align*}

\textbf{Step $\bm{4_0}$:} The protocol transfers whatever is left in the Transmission Line back to Alice's Laboratory by applying the operator $a^\dagger_{(\rm{A})}a_{(\rm{Tr})}$. In this scenario that leads to:
\begin{align*}
 & a^\dagger_{(\rm{A})}a_{(\rm{Tr})} \ket{0}^{(\rm{A})}\ket{0}^{(\rm{Tr})}\ket{1}^{(\rm{B})}  = 0 .
\end{align*}

\textbf{Step $\bm{5_0}$:} Alice applies a number operator, $a^\dagger_{(\rm{A})}a_{(\rm{A})}$, to her state and notes down the outcome. She will find that there is no particle in her domain. Bob empties his laboratory.

\textit{(b) The $1$-Bit Process:}

If Bob instead wishes to transmit a $1$-bit to Alice, the step after \textbf{Step $\bm{2}$} is instead given by:

\textbf{Step $\bm{3_1}$:} Bob inserts detectors in his laboratory i.e. in the upper IFM path. This causes collapse of the parts of the wavefunction that enter Bob's Laboratory and disables the self-interference of the interrogating particle. 
\begin{align*}
& a^\dagger_{(\rm{Tr})} \ket{0}^{(\rm{A})}\ket{0}^{(\rm{Tr})}\ket{0}^{(\rm{B})} \xrightarrow{BS_1}  \\ & \big[  \cos{(\theta)}a^{\dagger}_{(\rm{Tr})} + i\sin{(\theta)}a^{\dagger}_{(\rm{B})}\big]  \ket{0}^{(\rm{A})}\ket{0}^{(\rm{Tr})}\ket{0}^{(\rm{B})} \\ 
& = \cos{(\theta)}\ket{0}^{(\rm{A})}\ket{1}^{(\rm{Tr})}\ket{0}^{(\rm{B})} + i\sin{(\theta)}\ket{0}^{(\rm{A})}\ket{0}^{(\rm{Tr})}\ket{1}^{(\rm{B})} \\
 & \rightarrow 
 \begin{cases}
    \ket{0}^{(\rm{A})}\ket{1}^{(\rm{Tr})}\ket{0}^{(\rm{B})} , & \text{with $P=\cos{(\theta)}^2$},\\
    \ket{0}^{(\rm{A})}\ket{0}^{(\rm{Tr})}\ket{1}^{(\rm{B})} \rightarrow collapse, & \text{otherwise}.
  \end{cases}
  \\
  & \rightarrow ... \\
  & \rightarrow 
  \begin{cases}
    \ket{0}^{(\rm{A})}\ket{1}^{(\rm{Tr})}\ket{0}^{(\rm{B})} , & \text{with $P=\cos{(\theta)}^{2N}$},\\
    \ket{0}^{(\rm{A})}\ket{0}^{(\rm{Tr})}\ket{1}^{(\rm{B})} \rightarrow collapse, & \text{otherwise}.
  \end{cases}
\end{align*}

\textbf{Step $\bm{4_1}$:} The protocol again transfers whatever is in the Transmission Line to Alice's laboratory. The evolution now becomes:
\begin{align*}
  & a^\dagger_{(\rm{A})}a_{(\rm{Tr})} \begin{cases}
    \ket{0}^{(\rm{A})}\ket{1}^{(\rm{Tr})}\ket{0}^{(\rm{B})} , & \text{with $P=\cos{(\theta)}^{2N}$},\\
    collapse, & \text{otherwise}.
  \end{cases} \\
  & =
  \begin{cases}
    \ket{1}^{(\rm{A})}\ket{0}^{(\rm{Tr})}\ket{0}^{(\rm{B})} , & \text{with $P=\cos{(\theta)}^{2N}$},\\
    collapse, & \text{otherwise}.
  \end{cases}
\end{align*}

\textbf{Step $\bm{5_1}$:}  Alice applies the number operator, $a^\dagger_{(\rm{A})}a_{(\rm{A})}$, to her state and Bob empties his laboratory. In this process, Alice will find one particle in her laboratory with probability $P=\cos{(\theta)}^{2N}$ and thus records a logical $1$:
\begin{align*}
  &  
  a^\dagger_{(\rm{A})}a_{(\rm{A})} \begin{cases}
    \ket{1}^{(\rm{A})}\ket{0}^{(\rm{Tr})}\ket{0}^{(\rm{B})} , & \text{with $P=\cos{(\theta)}^{2N}$},\\
    collapse, & \text{otherwise}.
  \end{cases}
  \\
  & =
  \begin{cases}
    1\ket{1}^{(\rm{A})}\ket{0}^{(\rm{Tr})}\ket{0}^{(\rm{B})} , & \text{with $P=\cos{(\theta)}^{2N}$},\\
    collapse, & \text{otherwise}.
  \end{cases}
\end{align*}
Note that $\lim_{N \rightarrow \infty} \cos{(\theta = \pi / 2N)}^{2N}=1$, such that the protocol always succeeds if the number of perfect beam-splitters approaches infinity. This is an optical manifestation of the quantum Zeno effect \cite{Degasperis74, Misra77, Kwiat99}: the evolution into Bob's Hilbert space is suppressed by his frequent measurements of infinitesimally small parts of the wavefunction.

\section{Numerical Demonstration}
\begin{figure}
\centering
\includegraphics[scale=0.168]{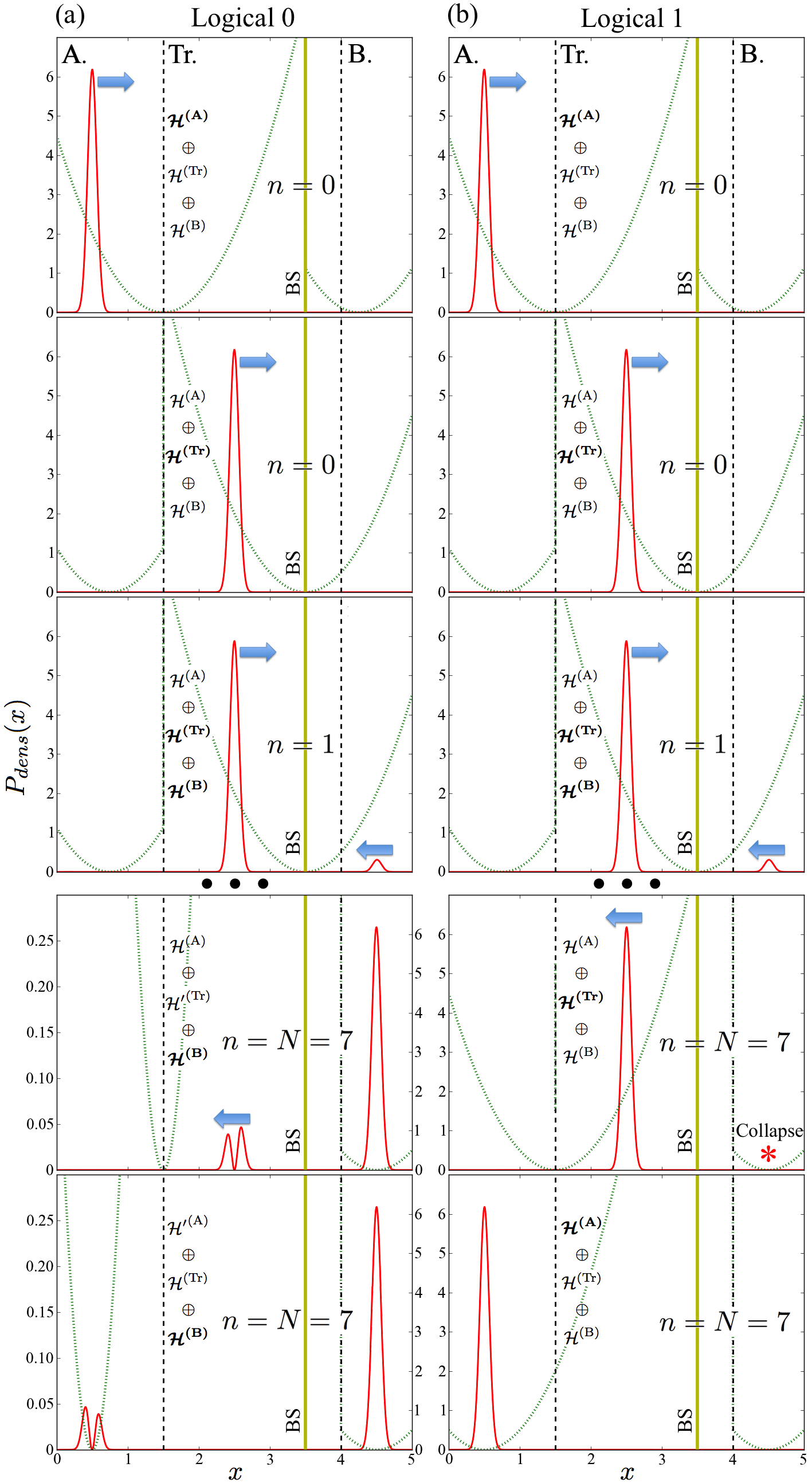}
\caption{(color online) This figure outlines the quantum evolution of the probability density distribution (solid red curve) of the 0-bit and 1-bit process in (a) and (b) respectively. The yellow line (at $x=3.5$) indicates the beam-splitter, the dotted green line shows the potential and the black dashed lines show the spatial divisions. The figure is further explained in the text.}
\label{fig:Demon}
\end{figure}

In order to evaluate the interaction-freeness of our protocol, we can ask ourselves: How does a quantum particle travel through the Hilbert space, $\mathcal{H}^{(\rm{A})} \oplus \mathcal{H}^{(\rm{Tr})} \oplus \mathcal{H}^{(\rm{B})}$, during our protocol? To answer this question we numerically solve the TDSE of a massive one-particle Hamiltonian that has been tailored to implement the above outlined scheme. The solution is obtained using an accelerated Staggered Leapfrog algorithm as in Ref. \cite{Owen14}. The wavefunction evolution is outlined in Fig. \ref{fig:Demon}.

The wavepacket is plotted at successive time frames (top to bottom). The particle is initialised in Alice's laboratory (A). It then falls into the Transmission Line (Tr) via a harmonic potential. The harmonic potential is shifted such that the particle hits the beam-splitter $N=7$ times (indicated by $n=1,\dotso,7$ in the figure). After each time, parts of the wavepacket enter Bob's laboratory. The Transmission Line is then emptied into Alice’s laboratory. In (b) Bob implements wavefunction collapse in his laboratory (B) after each beam-splitter interaction, whilst in (a), he does not. The last two frames in (a) have the Transmission Line and Alice's Laboratory magnified to show the failure probability density ($\sim$0.95 \%) of the protocol. This failure probability is due to errors in the beam-splitter caused by excitations into higher energy states of the harmonic wavepacket. The Hilbert space is written out on each frame and bold fonts denote the parts of the Hilbert space that are actively occupied at the specific frame. Primes denote weak Hilbert space occupation only caused by errors. \ref{fig:Demon}(b) shows the successful generation of a 1-bit event where the particle does not collapse into Bob's Laboratory. The probability of this is $\sim$70 \%.

\begin{figure}
\centering
\includegraphics[scale=0.18]{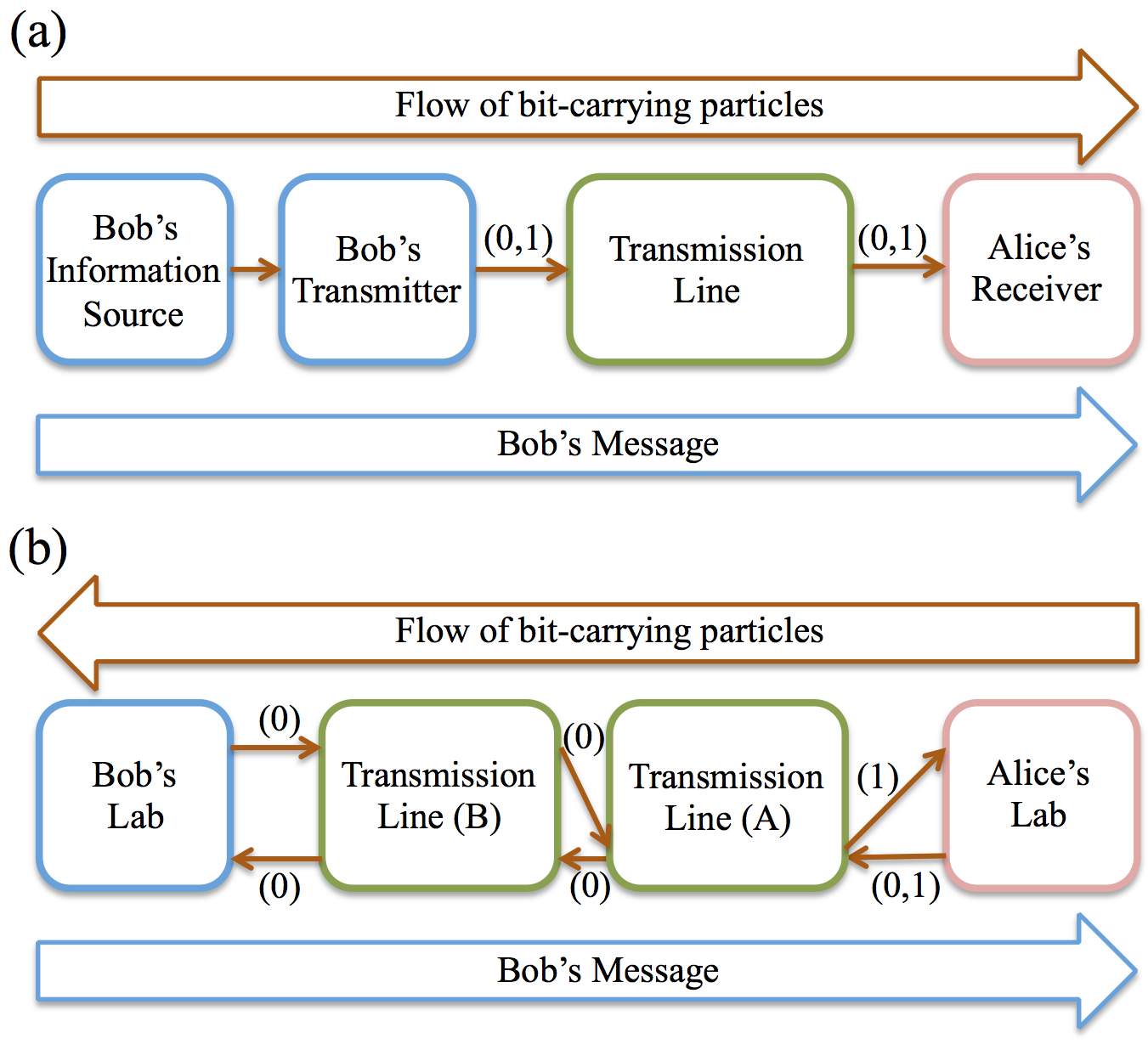}
\caption{(color online) The flow of particles and information in (a) classical communication schemes and in (b) weak-trace-free quantum counterfactual communication. The two parts of the Transmission Line are separated by a beam-splitter. In both cases information flows from Bob to Alice.}
\label{fig:Information}
\end{figure}

It becomes evident that, unless an error occurs, the wavefunction never evolves from Bob's space into Alice's. In the scenario of perfect channels and beam-splitters, weak measurements at any part of Bob's Laboratory will leave a measurable impact on the particles used in the $0$-bit process. However, for weak measurements, such particles still end up at Bob's Laboratory with a probability approaching unity. We thus conclude that the protocol is fully interaction-free from Alice's perspective. Furthermore, we coin the term ``Weak-Trace-Free Quantum Counterfactual Communication" to describe this phenomenon. A summary is given in FIG. \ref{fig:Information}.

\color{black}
We wish to highlight the fact that the previous attempt to realise counterfactual communication by Salih et al. \cite{Salih13} aimed at excluding any particles travelling between Alice and Bob. Our scheme does not do that. Particles do travel from Alice to Bob. However, Bob's message and Alice's particles are counterpropagating. We use Penrose's original definition (that counterfactuals are ``things that might have happened, although they did not in fact happen") \cite{Penrose94} and thus conclude that: from the receiver's perspective our protocol should be considered as fully counterfactual, as no particles actually traveled to it from the message source.
\color{black}

\section{Errors and Violations of Interaction-Freeness}
\begin{figure}
\centering
\includegraphics[scale=0.105]{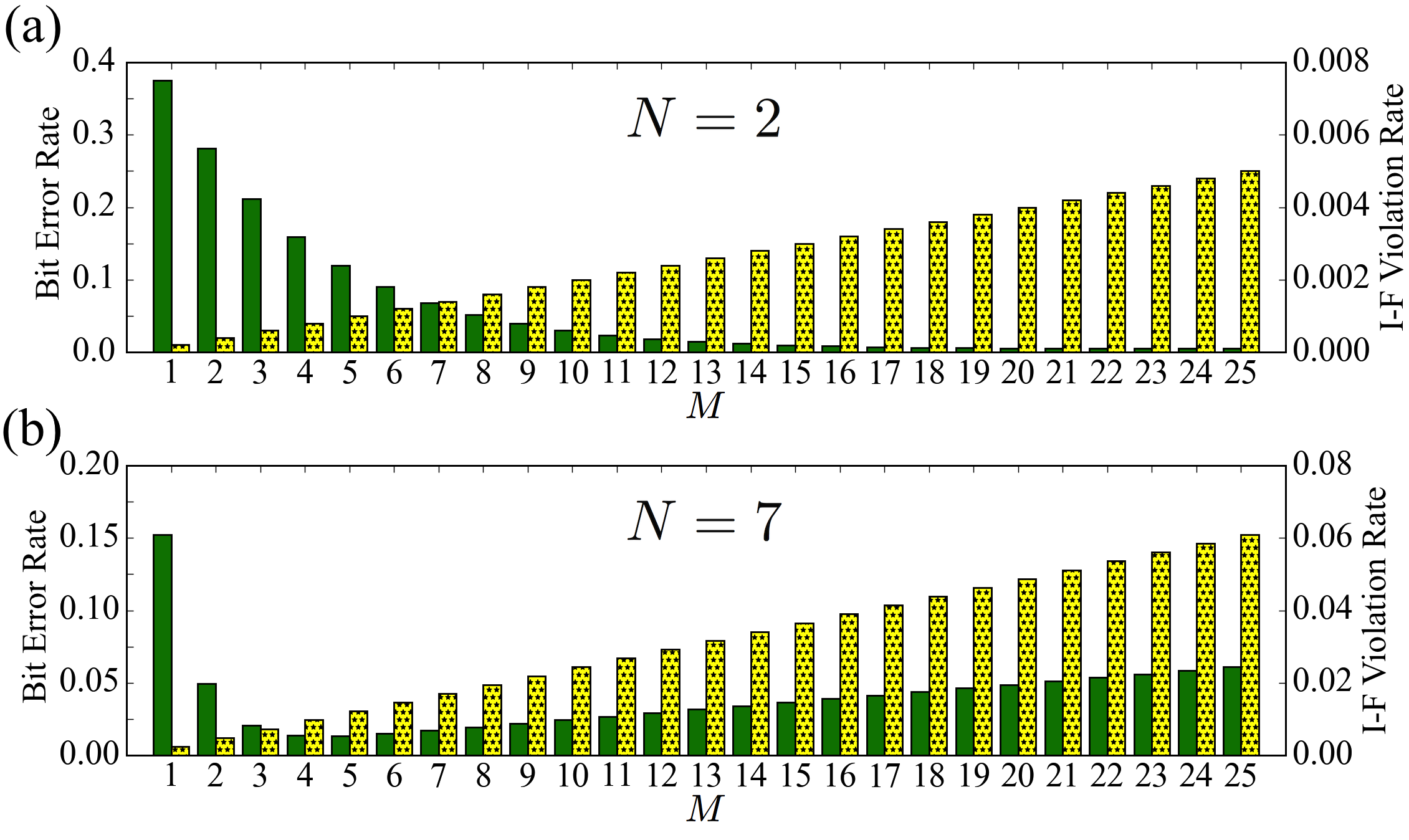}
\caption{(color online) Bit error rate (solid) and interaction-free violation rate (dotted) as functions of the process encoding number, $M$. The beam-splitter angle, $\theta = \pi / 2N$, was given a standard deviation of $\sigma_\theta = 0.01$ rad. (a) and (b) shows simulations of devices with $N=2$ and $N=7$ beam-splitters respectively.}
\label{fig:Errorz}
\end{figure}

It is experimentally challenging to stack a large number of beam-splitters. Furthermore, these beam-splitters naturally suffer from uncertainties in the unitary evolution. Hence, we now address the issue of the failure probabilities of the  $0$-bit and  $1$-bit processes: $P^{0}_{fail}$ and $P^{1}_{fail}$. For reasonable values of $N$ and high-fidelity beam-splitters, the nature of these probabilities causes the failure rate of the $1$-bit to be substantial and that $P^{0}_{fail} < P^{1}_{fail}$. However, we suggest an encoding such that a detection of one or more particles in Alice's Laboratory, out of $M$ processes, would constitute a logical $1$. The logical $0$ would be the scenario of no detections. The respective failure probabilities will then change such that $P^{0}_{fail,M} = (P^{0}_{fail}) \times M$ and $P^{1}_{fail,M} = (P^{1}_{fail})^M$. Whilst the failure probability of the logical $0$-process increases with increasing $M$, that of the logical $1$-process falls. Both failures generate bit errors, however, only the logical $0$-process failure generates a violation to the interaction-freeness of the protocol. Their respective significances can easily be tuned by $M$ as shown in FIG. \ref{fig:Errorz}.

We use Monte-Carlo simulations to explore the relations between the bit error rate and the interaction-free violation rate as functions of the process number, $M$, in devices with beam-splitters of non-perfect values of $\theta$. FIG. \ref{fig:Errorz} shows simulations with $10^9$ logical bit events. For low values of $N$, the average bit error is exponentially dependent on $P^{1}_{fail}$ for a significant number of $M$-values. When more beam-splitters are used, and $N$ is larger, the average bit error quickly becomes linearly dominated by $P^{0}_{fail}$  with increasing $M$. It is clear that if high-fidelity beam-splitters and quantum channels are available, even small values of $N$ allow for an effective reduction of the bit errors of the protocol, whilst still keeping the interaction-free violations small.

\section{Concluding Remarks}

In this paper we have outlined a weak-trace-free quantum counterfactual communication protocol that contradicts the classical perception of communication \cite{Shannon48}, by enabling the travel of information from Bob to Alice without any wavefunction travelling from Bob to Alice. Our protocol builds on interaction-free measurement devices \cite{Elitzur93, Kwiat95} and by numerically solving the Schr\"odinger equation we have demonstrated how it is realistically implementable with just a few beam-splitters. The protocol does not utilise nested MZIs as in previous \cite{Salih13} controversial suggestions for counterfactual communication. Numerical simulations show that---in the limit of perfect beam-splitters---our protocol does not have even infinitesimal parts of the wavefunction travelling from Bob's Laboratory to Alice's. Hence, it is immune to the weak-trace criticism of previous protocols \cite{Vaidman13}. Furthermore, whilst a substantial fraction of the individual $1$-bit processes might fail, we show how the logical bits can be redefined in terms of many processes such that the failure probability is only limited by the fidelity of the quantum channels and the unitary operations of the beam-splitters. The protocol is well within the realisable scope of quantum optics and we highly recommend experimental groups to pursue our work.

\bigbreak

The authors would like to express their gratitude towards Edmund Owen and Jacek Mosakowski for helpful discussions. This work was supported by the EPSRC and the Cambridge Laboratory of Hitachi Limited via Project for Developing Innovation Systems of the MEXT in Japan.

\bibliography{QuTel}

\end{document}